# Magnetotransport studies of Superconducting $Pr_4Fe_2As_2Te_{1-x}O_4$

A. Pisoni, P. Szirmai, S. Katrych, B. Náfrádi, R. Gaál, J. Karpinski, and L. Forró

*Institute of Physics, EPFL, CH-1015 Lausanne, Switzerland*

**Abstract**

We report a detailed study of the electrical transport properties of single crystals of $Pr_4Fe_2As_2Te_{1-x}O_4$, a recently discovered iron-based superconductor. Resistivity, Hall effect and magnetoresistance are measured in a broad temperature range revealing the role of electrons as dominant charge carriers. The significant temperature dependence of the Hall coefficient and the violation of Kohler's law indicate multiband effects in this compound. The upper critical field and the magnetic anisotropy are investigated in fields up to 16 T, applied parallel and perpendicular to the crystallographic *c*-axis. Hydrostatic pressure up to 2 GPa linearly increases the critical temperature and the resistivity residual ratio. A simple two-band model is used to describe the transport and magnetic properties of $Pr_4Fe_2As_2Te_{1-x}O_4$. The model can successfully explain the strongly temperature dependent negative Hall coefficient and the high magnetic anisotropy assuming that the mobility of electrons is higher than that of holes.

## I. INTRODUCTION

Eight years after their discovery, iron-based superconductors still represent a topic of vivid interest [1]. Many different compounds have been synthetized and studied with the hope of shedding light on the mechanism of high-temperature superconductivity in this class of materials [2,3]. In this quest, our group recently reported the synthesis and the characterization of a novel oxypnictide structure, $Ln_4Fe_2As_2Te_{1-x}O_4$ (*Ln*=Pr, Sm or Gd), that displays $T_c$= 25 K without fluorine doping and critical temperatures up to 46 K upon appropriate rare earth substitution and F-doping [4,5]. The intriguing aspect of this compound is the appearance of superconductivity even without any oxygen vacancy. The tellurium deficiency, naturally present in the crystal structure, appears to have a key role in the emergence of superconductivity in $Ln_4Fe_2As_2Te_{1-x}O_4$.

The characterization of the normal-state properties, like the Hall effect and magnetoresistance, are very important in understanding the carrier scattering processes occurring in a material. In cuprate superconductors, for example, the observation of non-linear temperature dependent resistivity and strongly temperature dependent Hall coefficients suggested the presence of unusual scattering mechanisms, beyond the typical electron-phonon interaction [6]. The upper critical field and the magnetic field anisotropy are interesting parameters, relevant for applications that can reveal multiband effects and give information about the pair-breaking mechanism. Hydrostatic pressure is also a remarkable and versatile tool for the investigation of the phase diagram of a material. By altering the interatomic distance, pressure can affect the electronic, magnetic and structural properties with the possibility of finding unique phenomena. Many studies have been performed on iron-based superconductors showing that pressure can greatly affect superconductivity in these compounds [3].

Here we present a systematic study of the Hall effect, magnetoresistance, upper critical field and pressure dependent resistivity of $Pr_4Fe_2As_2Te_{0.88}O_4$ in a broad range of temperature, magnetic field and pressure.

## II. EXPERIMENTAL DETAILS

Single crystals of $Pr_4Fe_2As_2Te_{0.88}O_4$ were grown at high pressure and at high temperature using a cubic multi-anvil system. A detailed description of the crystal growth and their characterization can be found elsewhere [4,5]. Focused ion beam (FIB) technique was employed to shape the crystals into a standard six-contact Hall bar configuration (inset Fig. 1(a)) and to deposit platinum leads on the samples' surface. In this way both longitudinal ($\rho_{xx}$) and transversal ($\rho_{xy}$) resistivity were measured at the same time on the same sample. The thickness of the crystals was accurately measured by SEM and it was typically between 8 and 10 $\mu$m. Measurement of magnetoresistance and Hall resistance were performed sweeping the magnetic field up to 10 T, at fixed temperatures down to 30 K. In order to evaluate the upper critical fields, $\rho_{xx}(T)$ was measured down to 1.7 K in different constant magnetic fields up to 16 T, applied both parallel and perpendicular to the FeAs-layers. The temperature dependence of $\rho_{xx}$ was also measured under pressure up to 2 GPa provided by a piston-cylinder cell. Daphne oil 7373 was employed as pressure transmitting medium to ensure completely hydrostatic conditions during pressurisation at room temperature in the pressure range investigated, as confirmed in some other works [7,8]. According to ref. [8], with this fluid, no drastic pressure drop has been observed upon cooling, which means that pressure remains quasi hydrostatic during the whole experiment. The pressure was determined by the superconducting transition temperature of a lead pressure gauge. Resistivity measurements were performed in a standard four-point configuration using a delta-mode technique to eliminate the thermoelectric voltages. In our analysis of the upper critical field and the effect of pressure on resistivity we define the onset $T_c$ at the crossing point of two extrapolated lines: one drawn through the resistivity curve in the normal state just above $T_c$, and the other through the steepest part of the resistivity curve in the superconducting state. The midpoint $T_c$ is determined as the temperature at which the resistivity is 50% of its value at the onset $T_c$. The zero-point $T_c$ is defined at the zero-resistivity point (i.e., when the measured $\rho_{xx}(T)$ drops below our experimental sensitivity). The graphical representation of these three different critical temperatures is provided in Fig. 4(b). From these three temperatures, we calculate the transition width as the difference between the onset and zero-point $T_c$.

## III. RESULTS AND DISCUSSION

### A. Magnetotransport

Fig. 1(a) presents the longitudinal resistivity as a function of temperature of a single crystal of $Pr_4Fe_2As_2Te_{0.88}O_4$ shaped by FIB (see Fig. 1(a) inset). The resistivity displays a non-linear variation in temperature with a substantial reduction of slope as the temperature increases. Such a saturation phenomenon was previously reported and extensively studied in many transition-metal compounds [9,10], Chevrel phases [11] and A15 superconductors [12,13]. It was interpreted in terms of the Mott-Ioffe-Regel (MIR) criterion, which states that resistivity saturates when the electron mean free path becomes comparable to the interatomic distance [14,15]. In A15 compounds resistivity is experimentally found to approach a limiting value between 100 and 300 µΩcm, in good agreement with the MIR prediction [15,16]. However, in $Pr_4Fe_2As_2Te_{0.88}O_4$ saturation occurs at much larger value ($\rho_{xx}(300K) = 1.3\,\text{m}\Omega\text{cm}$). Similar high saturation values were already observed in cuprate superconductors [17,18], alkali-doped fullerenes [19] and other iron-based superconductors [20,21]. In fact, the semi-classical Boltzmann theory on which the MIR criterion is based fails when one deals with strongly correlated multiband systems where the inter-band transitions become important [22-24]. Gunnarsson *et al.* demonstrated that the use of a more appropriate theoretical model allows resistivity to saturate at much higher values than predicted by the classical MIR criterion [22]. The saturation of the resistivity of K- and Ru-doped $BaFe_2As_2$, LiFeAs and $SrFe_2As_2$ compounds has been successfully explained in terms of a multiband scenario by Gobulov *et al.* [25]. They employed a two-band Eliashberg model where a strong disparity between the relaxation times and the coupling constants exist in different bands [25]. We therefore assume that this multiband model could explain as well the saturation of $\rho_{xx}(T)$ in $Pr_4Fe_2As_2Te_{0.88}O_4$. Indeed, several studies of angle-resolved photoemission spectroscopy (ARPES) have confirmed that iron-based superconductors have a rather complex multiband structure where five Fe $3d$ bands contribute to the Fermi surface [26].

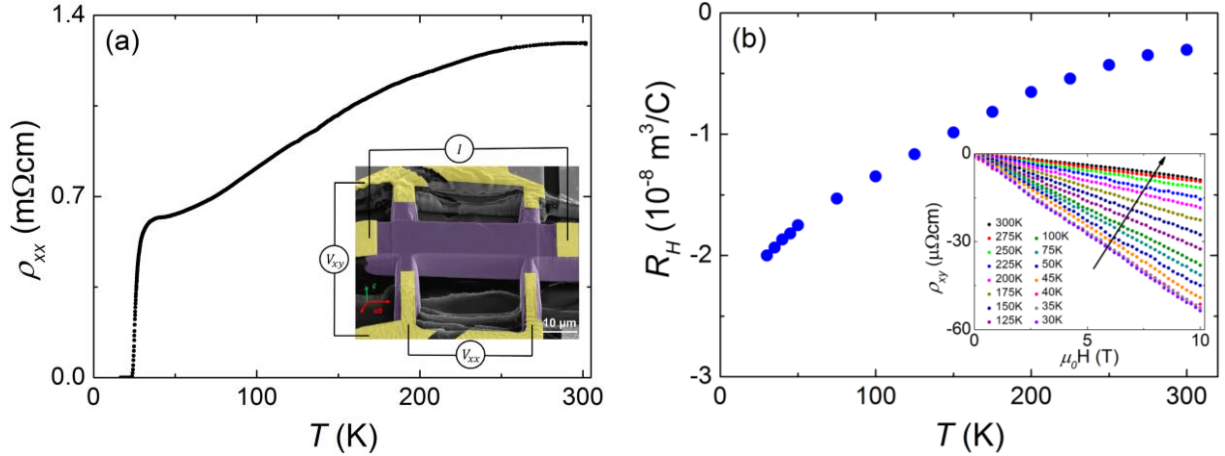

FIG. 1. (Color online) (a) Longitudinal electrical resistivity of the FIB-shaped $Pr_4Fe_2As_2Te_{0.88}O_4$ single crystal. The inset presents the shaped sample used for $\rho_{xx}$, $\rho_{xy}$ and Hall effect measurement. The sample is highlighted in violet and the platinum contacts in yellow. The current injection and the longitudinal and transversal voltages detection are presented by a circuit diagram. The crystallographic directions are also indicated by red arrows (*ab* plane) and a green arrow (*c* axis) at the bottom left of the inset. (b) Temperature dependent Hall coefficient of $Pr_4Fe_2As_2Te_{0.88}O_4$. The inset shows the transversal resistivity measured at different constant temperatures in magnetic field up to 10 T applied perpendicular to the sample's surface. The black arrow displays the increasing temperature according to the values reported in the legend.

The measured Hall coefficient ($R_H = \rho_{xy}/H$) of $Pr_4Fe_2As_2Te_{0.88}O_4$ is plotted as a function of temperature in Fig. 1(b). The inset in Fig.1 (b) presents the magnetic field dependence of the Hall resistivity ($\rho_{xy}$) at different temperatures. The transversal resistivity was measured as $\rho_{xy} = \left[\rho_{xy}(+H) - \rho_{xy}(-H)\right]/2$, to eliminate the effect of misaligned Hall electrodes. A linear dependence of $\rho_{xy}$ is observed from room temperature down to 30 K in different magnetic fields up to 10 T. This allows the unambiguous determination of the Hall coefficient by $R_H = \rho_{xy}/H$. The Hall coefficient is negative in the entire temperature range indicating that electrons are the dominant charge carriers. This result is in agreement with the data from structure refinement that confirm the presence of tellurium vacancies in the crystal structure [5]. These vacancies are responsible for electron doping in the $Fe_2As_2$ layers. However, the strong temperature dependence of $R_H$ suggests that electron and hole-like bands contribute to the charge transport and their compensation shapes the form of $R_H(T)$. In a single-band model $R_H = 1/qn$, where $q$ is the carrier charge and $n$ is the carrier density. In this case $R_H$ is almost *T*-independent. By contrast, the Hall coefficient in a two-band system consisting of electron and hole bands is given by:

$$R_H = \frac{1}{|q|} \frac{n_h \mu_h^2 - n_e \mu_e^2}{(n_h \mu_h + n_h \mu_h)^2} \quad, \tag{1}$$

where $n_h$ and $n_e$ are the hole and electron density, respectively, and $\mu_h$ and $\mu_e$ are the hole and the electron mobility, respectively. Equation (1) can successfully describe the temperature dependence of $R_H$ as the balance between the hole and electron bands is changed by *T*-dependent mobilities if the charges respond differently to phonons or spin fluctuations [27,28]. Other iron-based superconductors, like GdFeAsO$_{1-x}$F$_x$ [29] and SmFeAsO$_{1-x}$F$_x$ [30] for different fluorine doping, also show negative $R_H$ with comparable absolute values and a similar temperature dependence.

The magnetoresistance (MR) can provide important information about the electronic scattering processes and the details of the Fermi surface. It can help to deconvolute the temperature dependence of the electron and hole mobilities. The magnetoresistance was defined as $\Delta\rho = \rho_{xx}(H) - \rho_{xx}$, where $\rho_{xx}(H)$ is the longitudinal resistivity at a transversal magnetic field *H* and $\rho_{xx}$ is the longitudinal resistivity at zero-field. The inset to Fig. 2 shows the field dependence of the MR at different temperatures. The absolute value of the MR is rather low (2% at 30 K and at 10 T) and comparable to that of some fluorine-doped *Ln*FeAsO samples [27,31,32]. In the framework of semi-classical transport theory, Kohler's law [33] predicts that, for a system with a single isotropic relaxation time and a symmetric Fermi surface, the MR at different temperatures is scaled by the expression:

$$\frac{\Delta\rho}{\rho_{xx}} = \frac{\rho_{xx}(H) - \rho_{xx}}{\rho_{xx}} = f\left(H/\rho_{xx}\right) \quad, \tag{2}$$

where $f\left(H/\rho_{xx}\right)$ represents an arbitrary function of the magnetic field for the particular material, irrespective of temperature. Figure 2 presents the Kohler's plot for our compound. Evidently, the different curves do not overlap to form a single universal one, meaning that Kohler's law is not obeyed. The violation of Kohler's law was already observed in other multiband superconductors like in MgB$_2$ [34] and NdFeAsO$_{1-x}$F$_x$ [27]. The strong temperature dependent Hall coefficient and the violation of Kohler's law demonstrate that different relaxation times exist in different bands of Pr$_4$Fe$_2$As$_2$Te$_{1-x}$O$_4$ in the normal-state.

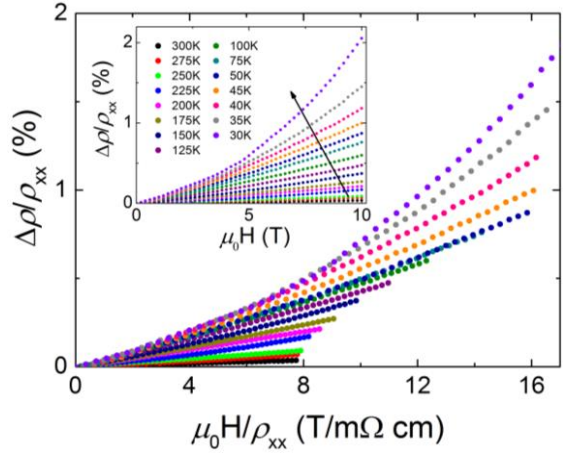

FIG. 2. (Color online) Kohler's plot evidencing the violation of Kohler's law as explained in the text. The inset shows the field dependence of magnetoresistance at different temperatures. The magnetoresistance is very low and decreases with increasing temperature. The black arrow displays the increasing temperature according to the values reported in the legend.

From the measurement of resistivity, magnetoresistance and the Hall effect is possible to determine the hole and electron mobilities $\mu_h$ and $\mu_e$ using a simple two-band model, with the assumption of equal electron and hole carrier density ($n = n_e = n_h$) [35]. In this case:

$$\rho_{xx} = \frac{1}{\sigma_e + \sigma_h} \quad , \tag{3}$$

$$R_H = R_e \frac{\sigma_h - \sigma_e}{\sigma_h + \sigma_e} \quad , \tag{4}$$

$$\frac{\Delta \rho}{\rho_{xx}} = R_e^2 \sigma_e \sigma_h H^2 \quad , \tag{5}$$

where $\sigma_{e,h} = nq\mu_{e,h}$ is the electron (or hole) conductivity, $R_e = 1/qn$ and $q$ is the absolute value of the electron charge. From these equations one finds that:

$$\frac{\Delta \rho}{\rho_{xx}} = \frac{1}{4}(R_e^2 - R_H^2)\left(\frac{H}{\rho_{xx}}\right)^2 \quad . \tag{6}$$

As explained by Albenque [35], Eq. (6) could be wrongly taken as an indication that Kohler's law is obeyed in a multiband system (with $n_e = n_h$) if the term $(R_e^2 - R_H^2)$ is weakly temperature dependent. This can happen if $R_e \gg R_H$ or if $R_H$ is nearly constant in $T$. Therefore, in such a case, the fulfilment of Kohler's rule should be considered with caution.

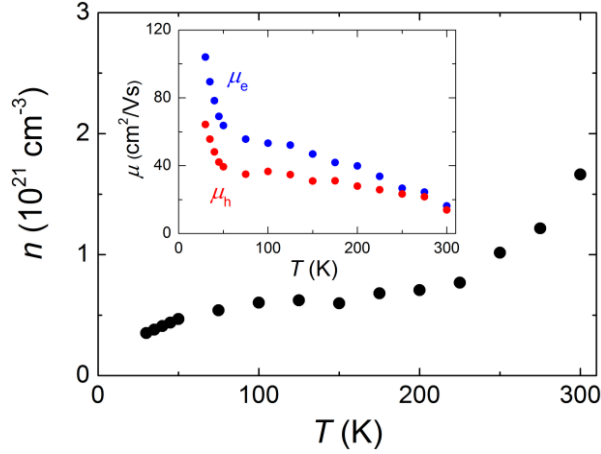

FIG. 3. (Color online) Temperature dependence of the carrier concentration evaluated from the compensated two-band model as explained in the text. The inset shows the calculated electron (blue points) and hole (red points) mobilities.

Figure 3 shows the temperature dependence of the charge carrier density ($n$) as estimated from Eq. (4)-(7). The value of $n$ at room temperature ($1.7 \times 10^{21}$ cm$^{-3}$) is quite low and is comparable to that reported in under-doped LaFeAsO$_{1-x}$F$_x$ [36]. The overall increase of $n$ with temperature is in contrast to the almost temperature independent behavior observed in some optimally doped pnictide superconductors [30]. From 300 K down to 230 K $n$ is strongly reduced. In this same $T$- range $\rho_{xx}$ displays saturation. Between 100 K and 200 K the carrier density remains almost constant but then decreases fast below 50 K. The inset in Fig. 3 presents the calculated electron (blue points) and hole (red points) mobilities. The hole and electron mobilities have a similar $T$ dependence. At high temperatures the difference between $\mu_e$ and $\mu_h$ is small but it increases with decreasing temperature. At low $T$, $\mu_e$ increases much faster than $\mu_h$, resulting in a more negative $R_H$. The sharp increase of both mobilities for $T < 50$ K is probably related to the freezing out of phonon modes which strongly scatter both electrons and holes. The non-linear temperature dependence of $n$, $\mu_h$ and $\mu_e$ accounts for that of the Hall coefficient. From the calculated mobilities we can extract an approximate value of the electron and hole scattering rate, $\left(\frac{m^*}{m_0}\right)\left(\frac{1}{\tau}\right)$ where $m^*$ is the effective mass, $m_0$ the electron mass and $\tau$ the relaxation time [35]. At 300 K we estimate $\left(\frac{m^*}{m_0}\right)\left(\frac{1}{\tau}\right) \simeq 10^{14} \text{s}^{-1}$ for both electrons and holes. The scattering rate decreases at 30 K to $1.7 \times 10^{13} \text{s}^{-1}$ and $2.7 \times 10^{13} \text{s}^{-1}$ for electrons and holes, respectively. These values appear in very good agreement with those reported in other iron-based superconductors like $BaFe_{2-x}Ru_xAs_2$ [35], LiFeAs [37] and $SmFeAsO_{1-x}F_x$ [38].

## B. Upper critical field

The single crystal shaped by FIB and used for the Hall effect measurement was also employed to extract the value of the upper critical field ($H_{c2}$) of $Pr_4Fe_2As_2Te_{1-x}O_4$. The longitudinal resistivity was measured as a function of temperature at different constant magnetic fields up to 16 T. The results for magnetic field oriented parallel to the crystallographic $c$-axis ($H \parallel c$) and perpendicular to it ($H \parallel ab$) are reported in Fig. 4(a) and 4(b), respectively. In zero-field the midpoint critical temperature of our crystals is $T_c = 25.8$ K. With increasing magnetic field the width of the superconducting transition increases monotonously and $T_c$ is suppressed more when $H \parallel c$ than when $H \parallel ab$, in agreement with the general trend reported for layered superconductors [39,40]. For $H \parallel c$ the zero-resistance state could not be reached above 12 T down to 1.7 K. The behaviour of $\rho_{xx}(T)$ for $H \parallel c$ observed in $Pr_4Fe_2As_2Te_{0.88}O_4$ appears in contrast to what we recently reported in the isostructural $SmFeAsTe_{1-x}O_{4-y}F_y$. In that case a pronounced sharpening of the superconducting transition is observed upon increasing field for $H \parallel c$ [41].

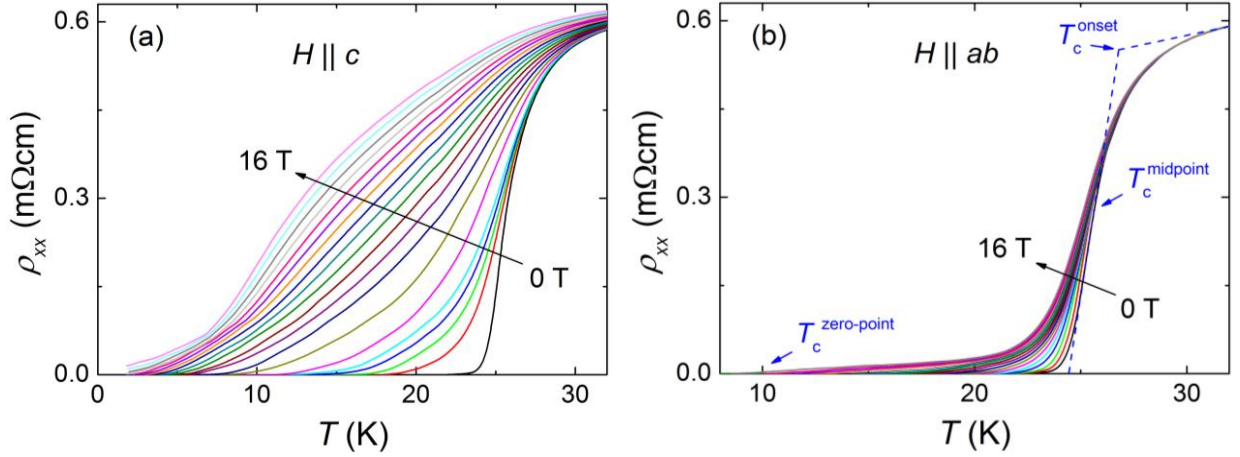

FIG. 4. (Color online) Temperature and magnetic field dependence of the resistivity of $Pr_4Fe_2As_2Te_{0.88}O_4$ measured with fields applied (a) perpendicular to the $Fe_2As_2$ layers ($H \parallel c$) and (b) parallel to them ($H \parallel ab$), the dashed blue line represents the criteria used to evaluate the three different critical temperatures as stated in the text. The values of the magnetic field are 0, 0.25, 0.5, 0.75 T and then from 1 to 16 T in 1 T steps, following the arrow in both panels.

Figure 5(a) presents the upper critical field of $Pr_4Fe_2As_2Te_{0.88}O_4$ as a function of temperature for $H \parallel c$ and $H \parallel ab$ defined at the midpoint $T_c$ (filled dots) and at the zero-resistivity point (empty dots), as schematically presented in Fig. 4(b). At the midpoint $T_c$ the in-plane ($H_{c2}^{ab}$) upper critical field shows a linear temperature dependence up to 16 T, while the out-of-plane ($H_{c2}^{c}$) upper critical field deviates from it above 10 T displaying a light convex curvature, as usually observed in 1111 compounds [39]. Due to the relatively low magnetic field applied, the extrapolation of the zero-temperature upper critical field ($H_{c2}(0)$) is particularly difficult and it can be only done in a very approximate way. To analyse our experimental results we evoke a two-band model which accounts for inter-band and intra-band scattering and that was successfully employed for $MgB_2$ and $NdFeAsO_{0.7}F_{0.3}$ [42-45]. According to this model $H_{c2}$, taking into account both orbital and paramagnetic pair breaking effects, can be written in the following parametric form [46]:

$$H_{c2} = \frac{2\Phi_0 k_B T_c t x}{\hbar D_0} \quad , \tag{7}$$

$$\ln t = -\frac{1}{2}\left[U_1(x) + U_2(x) + \frac{\lambda_0}{w}\right] + \text{sign}(w)\left\{\frac{1}{4}\left[U_1(x) - U_2(x) - \frac{\lambda_-}{w}\right]^2 + \frac{\lambda_{12}\lambda_{21}}{w^2}\right\}^{1/2} \quad , \tag{8}$$

$$U_{1,2}(x) = \text{Re}\left\{\psi\left[\frac{1}{2} + \left(i + \frac{D_{1,2}}{D_0}\right)x\right] - \psi\left(\frac{1}{2}\right)\right\} \quad , \tag{9}$$

$$\lambda_\pm = \lambda_{11} \pm \lambda_{22} \quad , \tag{10}$$

$$w = \lambda_{11}\lambda_{22} - \lambda_{12}\lambda_{21} \quad, \tag{11}$$

$$\lambda_0 = \left(\lambda_-^2 + 4\lambda_{12}\lambda_{21}\right)^{1/2} \quad, \tag{12}$$

where $t = T/T_c$, $\psi(x)$ is the digamma function, $D_1$ and $D_2$ are the diffusivity in band 1 and 2, $D_0 = \hbar/2m$ where $m$ is the electron mass, $k_B$ is the Boltzmann constant and $\Phi_0$ is the magnetic flux quantum. The constants $\lambda_{11}$ and $\lambda_{22}$ represent the intra-band pairing in bands 1 and 2 respectively, while $\lambda_{12}$ and $\lambda_{21}$ quantify the inter-band coupling. The parameter $x$ runs from 0 to $\infty$ as $T$ varies from $T_c$ to 0. If the magnetic field is inclined by an angle $\theta$ with respect to the $ab$ planes the diffusivities $D_{1,2}$ must be replaced by their angular dependence:

$$D_{1,2}(\theta) = (D_{1,2}^{(ab)2}\sin^2\theta + D_{1,2}^{(ab)}D_{1,2}^{(c)}\cos^2\theta)^{1/2} \quad, \tag{13}$$

where $D_{1,2}^{(ab)}$ and $D_{1,2}^{(c)}$ are the in-plane and $c$-axis values in bands 1 and 2 [45]. The values of $H_{c2}(0)$ for the two crystallographic directions are estimated using the best fit of Eq. (7)-(12) to the experimental data and extending the fits to lower temperatures. The values of $\lambda_{ij}$ and $D_{1,2}$ are allowed to vary to minimize the root-mean-square error. However, as it was already reported in NdFeAsO$_{0.7}$F$_{0.3}$ [45], the resulting fits are rather weakly sensitive to the particular choice of the coupling constants $\lambda_{ij}$ but they mostly depend on the ratio $D_2/D_1$. The blue solid lines in Fig. 5(a) and (b) represent the best fits to the upper critical field data, measured with the midpoint criterion, using $\lambda_{11} = \lambda_{22} = 0.5$, $\lambda_{12} = \lambda_{21} = 0.4$, $D_2^{(ab)}/D_1^{(ab)} = 0.15$, $D_2^{(c)}/D_1^{(c)} = 0.1$, $D_1^{(ab)} = 0.3D_0$ and $D_1^{(c)} = 3.2D_0$. The $w>0$ and $\lambda_{11}\lambda_{22} > \lambda_{12}\lambda_{21}$ indicate dominant intra-band coupling, while the low ratio $D_2/D_1$ suggests that the mobility of the charge carriers in one band is different from that in the other [45]. This appears in agreement with the results from the Hall effect measurement. As it can be seen in Fig. 5(a), the two-band model gives a very good description of the experimental results. We also tried to fit the data using a simple linear fit in temperature, as proposed by the Werthamer-Helfand-Hohenberg (WHH) theory [47]. The linear fits are presented as dashed pink lines in Fig. 5(a) and (b).

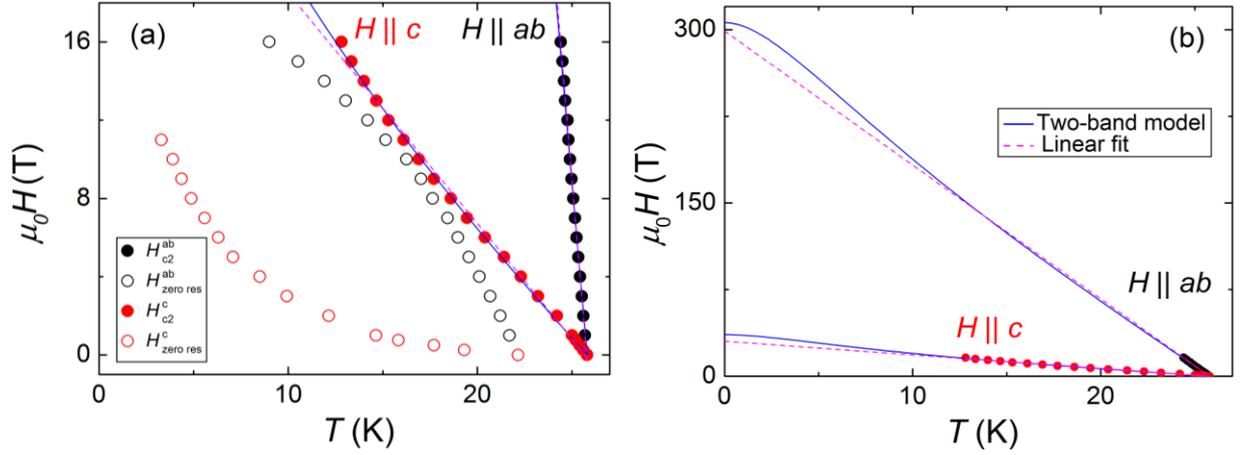

FIG. 5. (Color online) (a) Upper critical magnetic field as a function of temperature from resistivity measurements with $H \parallel ab$ (black points) and $H \parallel c$ (red points). Filled dots correspond to values of the upper critical field ($H_{c2}$) estimated from the midpoint of the resistive transitions while empty dots represent the fields evaluated at the zero-resistivity point. Blue dashed lines are the linear fits to the experimental data. The solid blue lines are fits to the two-band model while the pink dashed lines are linear fits to the data as explained in the text. (b) Low-temperature region of the two-band model and the linear fit to the magnetic field values measured with the midpoint criterion.

From the two-band model we obtain $\mu_0 H_{c2}^{ab}(0) = \Phi_0/(2\pi \xi_{ab} \xi_c) = 306$ T and $\mu_0 H_{c2}^{c}(0) = \Phi_0/(2\pi \xi_{ab}^2) = 36$ T, where $\xi_{ab}$ and $\xi_c$ are the in-plane and the out-of-plane coherence lengths. From the previous equations we deduce $\xi_{ab} = 30$ Å and $\xi_c = 3.6$ Å. The values of $H_{c2}(0)$ estimated by the two-band model are larger than those predicted by the WHH model, $H_{c2}(0) = -0.69 T_c \left( dH_{c2}/dT \right)_{T_c}$, that gives $\mu_0 H_{c2}^{ab}(0) = 200$ T and $\mu_0 H_{c2}^{ab}(0) = 20$ T. However, it's important to note that, as observed in other quasi-two dimensional superconductors such as organic, cuprate and Fe-based superconductors, the low-field temperature dependence of $H_{c2}$ is often a poor guide to the estimation of $H_{c2}(0)$ and its detailed behavior at low temperatures [48].

The magnetic field anisotropy ($\gamma_H = H_{c2}^{ab}/H_{c2}^{c}$), calculated using the midpoint criterion, is $\gamma_H \sim 8.5$ close to $T_c$. This value is slightly higher than that observed in many other FeAs-based superconductors [32,44,45,49,50] and it reflects the remarkable structural anisotropy of $Pr_4Fe_2As_2Te_{1-x}O_4$.

## C. Pressure effect

The $\rho_{xx}(T)$ of $Pr_4Fe_2As_2Te_{0.88}O_4$ measured at different hydrostatic pressures (*p*) up to 2 GPa is presented in Fig. 6(a). At high temperature resistivity approaches saturation for all the applied pressures, exactly as we recently described in $SmFeAsTe_{1-x}O_{4-y}F_y$ [41]. As pressure increases the slight up-turn observed in $\rho_{xx}(T)$ at ambient pressure just above $T_c$ is strongly reduced and it totally disappears at 2 GPa. The inset of Fig. 6(a) shows the pressure dependence of the extrapolated zero-temperature residual resistivity ($\rho_{xx}^0$) (black dots) and the room temperature resistivity (red dots) normalized to their value at atmospheric pressure. The blue lines in the inset of Fig. 6(a) are linear fits to the data. At 1 bar $\rho_{xx}(300K) = 1.3\,m\Omega cm$ and $\rho_{xx}^0 = 0.38$ mΩcm. With increasing pressure $\rho_{xx}(300K)$ decreases at a rate of ~18% /GPa reaching a value of 0.81 mΩcm at 2 GPa. We also observe a drastic decrease of $\rho_{xx}^0$ which falls almost linearly by more than 60% under pressure ($\rho_{xx}^0 = 0.14\,m\Omega cm$ at 2GPa). This result is very surprising because such a large suppression of $\rho_{xx}^0$ by pressure is usually observed in materials displaying collective modes, like charge [51] or spin density waves [52], and in heavy fermions systems where the increase of pressure is believed to reduce the scattering of charge carriers by magnetic fluctuations [53]. However, similar results were also obtained for superconducting $YBa_2Cu_3O_{7-\delta}$, where the out-of-plane residual resistivity was reduced up to 50 % in the same pressure range [54]. $\rho_{xx}(T)$ may contain both the in-plane ($\rho_{ab}$) and the out-of-plane ($\rho_c$) resistivity component due to extended in-plane defects which force the charge carriers for inter-plane transitions. We assume that the significant suppression of $\rho_{xx}^0$ in $Pr_4Fe_2As_2Te_{0.88}O_4$ can be primary ascribed to a considerable decrease of $\rho_c^0$, likely in the oxygen deficient $YBa_2Cu_3O_{7-\delta}$. In fact, the tellurium vacancies could be responsible for the strong drop of $\rho_{xx}^0$: the "voids" caused by Te deficiency in the spacing layers of $Pr_4Fe_2As_2Te_{0.88}O_4$ can be easily minimized by pressure, thus ensuring a better connectivity and charge conduction along the crystal *c*-axis. The strong suppression of $\rho_{xx}^0$ and of the resistivity up-turn above $T_c$ clearly demonstrates that pressure leads to an optimization of the crystal structure of $Pr_4Fe_2As_2Te_{0.88}O_4$.

Another possibility is that the strong pressure variation of the resistivity is due to the multiband character of the electronic structure of our compound. Pressure can influence band structure by increasing the bandwidth, which implies changes in the carrier effective masses, band overlaps and the electronic density of states (N(E)). A simple two-band model, as proposed by Sales *et al.* [55] to describe the transport properties of the Ba(Fe$_{1-x}$Co$_x$)$_2$As$_2$ superconductor, can explain the reduction of $\rho_{xx}(300K)$ with pressure causing an increase in electron and hole band overlap [52]. The *p*-dependence of the density of state at the Fermi level (N(E$_F$)) can also significantly affect the residual resistivity. For example, studies on Au-Pd alloys under pressure have shown that a small decrease of the lattice constant (from 7.6 to 7.4 a.u.) reduces $\rho_{xx}^0$ up to 25 % in Au$_{25}$Pd$_{75}$ [56]. Such a large reduction of $\rho_{xx}^0$ was instead not observed in Au-rich alloys [56]. This is because E$_F$ in Pd-rich alloys is close to the band edge and a little change in pressure provokes a strong reduction of N(E$_F$) [56]. In Au-rich alloys, however, E$_F$ is in a region where the density of states is almost constant and variations of pressure cause only small changes in $\rho_{xx}^0$ [56]. The different rate at which $\rho_{xx}(300K)$ and $\rho_{xx}^0$ decrease with pressure causes an increase of the residual resistance ratio $RRR = \rho(300K)/\rho(0K)$ from 2.4 at 0 GPa to 6.8 at 2 GPa. The variation of $T_c$ with pressure is presented in Fig. 6(b). The onset $T_c$, the midpoint $T_c$ and the zero-point $T_c$ are defined as explained in the introduction (See also Fig.4b). The critical temperature increases almost linearly at a rate of 0.5 K/GPa (dashed lines in Fig. 6(b)). Similarly to the case of cuprate superconductors such an increase of $T_c$ with pressure could be interpreted as an indication of sample in an under-doped state. Very recent *ab-initio* calculations have demonstrated that hydrostatic pressure decreases the lattice parameters of Pr$_4$Fe$_2$As$_2$Te$_{1-x}$O$_4$ [57]. Because of the conservation of the Fe-As bonding length a compression in the *ab* plane results in an increase of the pnictogen height ($h_{As}$). In this way $h_{As}$ of Pr$_4$Fe$_2$As$_2$Te$_{1-x}$O$_4$ can approach the optimum value of 1.38 Å, that was found to be related to a maximum $T_c$ in iron-based superconductors. [58].

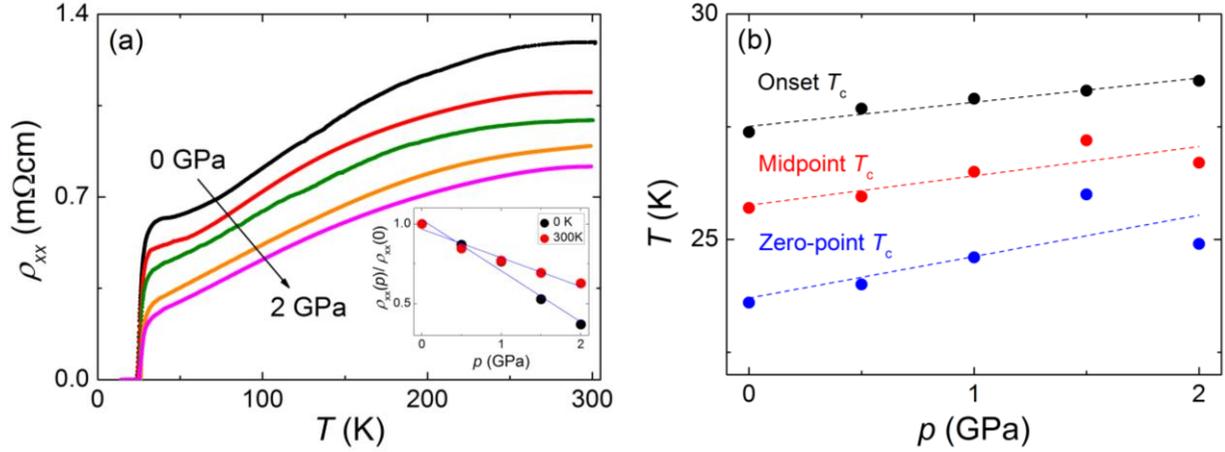

FIG. 6. (Color online) (a) Electrical resistivity versus temperature at different pressures up to 2 GPa shows the pressure dependence of the extrapolated zero-temperature residual resistivity (black dots) and the room temperature resistivity (red dots) normalized to their value at atmospheric pressure. The blue lines are linear fits to the data. (b) Critical temperatures evaluated with different criteria (see text) as a function of pressure. The dashed lines are linear fits to the data.

## IV. CONCLUSIONS

We measured the Hall coefficient, magnetoresistance, upper critical field and temperature dependent resistivity of $Pr_4Fe_2As_2Te_{0.88}O_4$ under different hydrostatic pressures. The Hall coefficient has a negative sign, indicating dominant electron-type charge carriers, and it is strongly temperature dependent. The value of magnetoresistance is very low (2% at 30 K and at 10 T) and it decreases with increasing temperature. The violation of the Kohler's law and the temperature dependent Hall coefficient are evidence of multiband effects in the normal-state of $Pr_4Fe_2As_2Te_{0.88}O_4$. A simple compensated two-band model is used to describe the resistivity, Hall effect and magnetoresistance data. The carrier density extracted by this model shows a relatively low value at room temperature that further decreases with lowering temperature. The calculated electron and hole mobilities increase in a similar way at low temperature. Electron mobility is higher than that of holes in the entire temperature range as testified by the negative Hall coefficient. The zero-temperature upper critical fields, extracted using a two-band model, are $\mu_0 H_{c2}^{ab}(0) \simeq 306$ T and $\mu_0 H_{c2}^{c}(0) \simeq 36$ T parallel and perpendicular to the *ab* planes, respectively. A relative high value of magnetic anisotropy near $T_c$ ($\gamma_H \sim 8.5$) reflects the large *c* lattice parameter of the unit cell of $Pr_4Fe_2As_2Te_{1-x}O_4$. The critical temperature and the *RRR* of $Pr_4Fe_2As_2Te_{0.88}O_4$ increase with hydrostatic pressure, in agreement with recent theoretical calculations. All these results demonstrate that the transport and magnetic properties of the $Pr_4Fe_2As_2Te_{1-x}O_4$ superconductor are significantly affected by its multiple band nature.

**ACKNOWLEDGEMENT**

This work was supported by the Swiss National Science Foundation (Project No. 140760, 144419 and 156012) and by the European Community FP7 Super-Iron Project.